# Music to *My* Ears: Neural modularity and flexibility differ in response to real-world music stimuli

**Melia E. Bonomo[1,2][†], Anthony K. Brandt[3][*][†], J. Todd Frazier[4], Christof Karmonik[4,5,6]**

[1] Department of Physics and Astronomy, Rice University, Houston, TX, USA
[2] Center for Theoretical Biological Physics, Rice University, Houston, TX, USA
[3] Shepherd School of Music, Rice University, Houston, TX, USA
[4] Center for Performing Arts Medicine, Houston Methodist Hospital, Houston, TX, USA
[5] MRI Core, Houston Methodist Research Institute, Houston, TX, USA
[6] Department of Radiology, Weill Cornell Medical College, New York, NY, USA

**\* Correspondence:**
Melia E. Bonomo
Present Address: Department of Bioengineering, Rice University, Houston, TX, USA
mbonomo@alumni.rice.edu

[†] These authors contributed equally to this work.

**Abstract**

Music listening involves many simultaneous neural operations, including auditory processing, working memory, temporal sequencing, pitch tracking, anticipation, reward, and emotion, and thus, a full investigation of music cognition would benefit from whole-brain analyses. Here, we quantify whole-brain activity while participants listen to a variety of music and speech auditory pieces using two network measures that are grounded in complex systems theory: modularity, which measures the degree to which brain regions are interacting in communities, and flexibility, which measures the rate that brain regions switch the communities to which they belong. In a music and brain connectivity study that is part of a larger clinical investigation into music listening and stroke recovery at Houston Methodist Hospital's Center for Performing Arts Medicine, functional magnetic resonance imaging (fMRI) was performed on healthy participants while they listened to self-selected music to which they felt a positive emotional attachment, as well as culturally familiar music (J.S. Bach), culturally unfamiliar music (Gagaku court music of medieval Japan), and several excerpts of speech. There was a marked contrast among the whole-brain networks during the different types of auditory pieces, in particular for the unfamiliar music. During the self-selected and Bach tracks, participants' whole-brain networks exhibited modular organization that was significantly coordinated with the network flexibility. Meanwhile, when the Gagaku music was played, this relationship between brain network modularity and flexibility largely disappeared. In addition, while the auditory cortex's flexibility during the self-selected piece was equivalent to that during Bach, it was more flexible during Gagaku. The results suggest that the modularity and flexibility measures of whole-brain activity have the potential to lead to new insights into the complex neural function that occurs during music perception of real-world songs.

# 1   Introduction

Visit any human habitation on earth and you are likely to find music woven into the fabric of life: in a meta-analysis of 315 societies, Mehr et al. (2019) found evidence of music in all of them. Yet although omnipresent as a cultural phenomenon, music itself—ranging from a Western orchestra to Aka polyphony, Tuvan throat singing, and Maori powhiri—is extremely diverse. As Trehub et al. (2015) write "Strictly speaking, there are no structural characteristics that have been identified in all known musical systems" (p. 2). Given this heterogeneity, many scientists distinguish between *musicality* and *music* (Honing et al., 2015). As Patel (2019) explains, *musicality* refers to "the set of mental capacities underlying basic musical behavior," whereas *music* "is a construct highly dependent on culture" (p. 460). *Music cognition* within an individual brain lies at the intersection of the two: it is dependent on *musicality* but shaped by *culture*.

Patel (2019) has written that "music cognition is not a unitary mental phenomenon and instead involves a collection of distinct and interacting mental processes" (p. 459). This includes auditory processing, working memory, temporal sequencing, pitch tracking, anticipation, reward, and emotion (Zatorre and Salimpoor, 2013). The complex combination of neural operations performed in the brain during music cognition necessitates analytic methods that take the functional activity of the whole-brain into account. Furthermore, the use of real-world musical stimuli is needed to begin to answer questions about the interaction between musicality and culture within the brain.

A novel functional magnetic resonance imaging (fMRI) study we have undertaken at Houston Methodist Hospital's Center for the Performing Arts Medicine addresses both the needs for a whole-brain analysis method and presentation of real-world stimuli. As part of a larger investigation of music's role in stroke rehabilitation, our goal was to compare the neurological responses between various musical and speech stimuli. Participants without prior musical training were asked to listen to long excerpts of six stimuli: a self-selected track for which they felt a positive emotional connection, examples of music that were culturally familiar (J.S. Bach) and unfamiliar (Gagaku court music of medieval Japan), emotional and unemotional English speech, and speech from a foreign language.

To quantify differences in brain response for the different auditory pieces, we focus on two whole-brain network measures: modularity and flexibility. Modularity has been widely applied to study brain networks (Sporns and Betzel, 2016) as it measures the degree to which brain regions can be grouped into modules based on their structural connections or functional network of interactions (Newman, 2006). Flexibility is the rate that brain regions change their module membership (Bassett et al., 2011), and therefore can measure how dynamic the network structure is while the brain performs a particular task. Importantly for the motivation of this work, modularity and flexibility are principles of design that are rooted in complex systems theory (Simon, 1962) and appear in diverse biological systems besides the brain (Bonomo, 2020). The degree of dynamic, modular structure in brain networks is associated with differences in cognitive performance under different task demands. In previous work with this dataset, in which we only looked at the static modularity, we found that those with higher modularity during the self-selected song exhibited the biggest change in modularity during the more novel auditory stimuli, while the familiar stimuli led to less perturbation of the network structure (Bonomo et al., 2020). Prior theory modeled the benefit of high modularity for performing fast, simple cognitive tasks and the benefit of low modularity for longer, more complex tasks (Chen and Deem, 2015), and experiments have demonstrated this dichotomous connection between performance and both resting-state (Yue et al., 2017) and task-based (Lebedev et al., 2018) modularity. The opposite relationship has been experimentally observed for flexibility, where low flexibility correlates with performance on simple tasks, and high flexibility correlates with performance on complex tasks (Ramos-Nuñez et al., 2017). Furthermore, there is a negative relationship between modularity and flexibility in resting-state fMRI data (Ramos-Nuñez et al., 2017). Here, we look at the modularity-flexibility relationship during task-based fMRI to study how the brain processes auditory pieces of varying familiarity.

For the musical pieces, we find that during the self-selected song and Bach, there is a significant negative correlation between a participant's whole-brain modularity and flexibility. This relationship largely disappears when the culturally unfamiliar Gagaku music is played. Furthermore, the auditory cortex is equally flexible during the self-selected piece and Bach, while it was more flexible during Gagaku. We hypothesize that the negative modularity-flexibility correlation may denote that the brain, as a complex system, is configured to efficiently process familiar stimuli, whereas it may be driven out of this configuration by highly novel stimuli that require more effort to process. Overall, our results suggest that the modularity and flexibility measures of whole-brain activity have the potential to lead to new insights into the complex neural function that occurs during music perception, in particular during real-world music stimuli.

## 2  Methods

We performed fMRI as 25 healthy adult participants actively listened to six excerpts of music and speech. The neuroimaging run for each auditory piece lasted 312 seconds. Further technical details about the participants and scans are found in the Extended Methods Section. The auditory pieces included a self-selected song (Self) and a playlist created by the researchers. The playlist consisted of auditory selections chosen for their cultural familiarity and unfamiliarity to the participants in the study (Bach and Gagaku, respectively), emotional speech from Charlie Chaplin in the film "The Great Dictator" (Chaplin), an unemotional newscast read by Walter Cronkite (Cronkite), and unfamiliar foreign speech from the South African Xhosa tribe (Xhosa).

We were particularly interested in seeing the contrasts in whole-brain activity during the different musical pieces. For the self-selected piece, participants were instructed to choose a song to which they felt a strong emotional attachment. For the culturally familiar music, we chose J.S. Bach's 2-part Invention in C-Major, BWV 772, a short piano work representative of traditional classical music originating in Europe during the common practice period (17th to early 20th centuries AD). For the culturally unfamiliar music, we selected a performance of Gagaku, the court music of Japan. Dating from the 8th–12th centuries AD, Gagaku is widely considered to be the oldest orchestral music in the world and one of the oldest unbroken musical traditions. However, in both sound and rhetoric, this aristocratic music is considered "remote" and "esoteric" (Tanaka and Koto, 2016, p. 18). It is particularly disorienting for naïve listeners given the unique instrumental techniques, including "glissandi, an accelerating repetition of the same note, an undulation of the notes, noises such as that of breathings [and] shouts, etc…" (Tamba, 1976, p. 8). The Gagaku track thus provided a strong contrast to the other musical selections. Indeed, by quantifying perceptual musical features, Bach and Self songs were more musically similar to each other than either was to Gagaku (see Extended Methods section)

To conduct a network analysis, we divided the brain into 84 anatomical regions (Brodmann areas, or BAs) and averaged the BOLD signal over all fMRI voxels in each region (see **Figure 1**). If two brain regions exhibited similar BOLD signal time series during the neuroimaging run, we inferred that these regions were working together to process the stimulus and drew a network link between them. The resulting functional activity network was then representative of how each auditory piece was processed by the whole brain. In our analysis, we focused on two measures to quantify the network structure: modularity (Newman, 2006), which gave us an overall summary of the brain network, and flexibility (Bassett et al., 2011), which gave us information about how dynamic the brain network was over time.

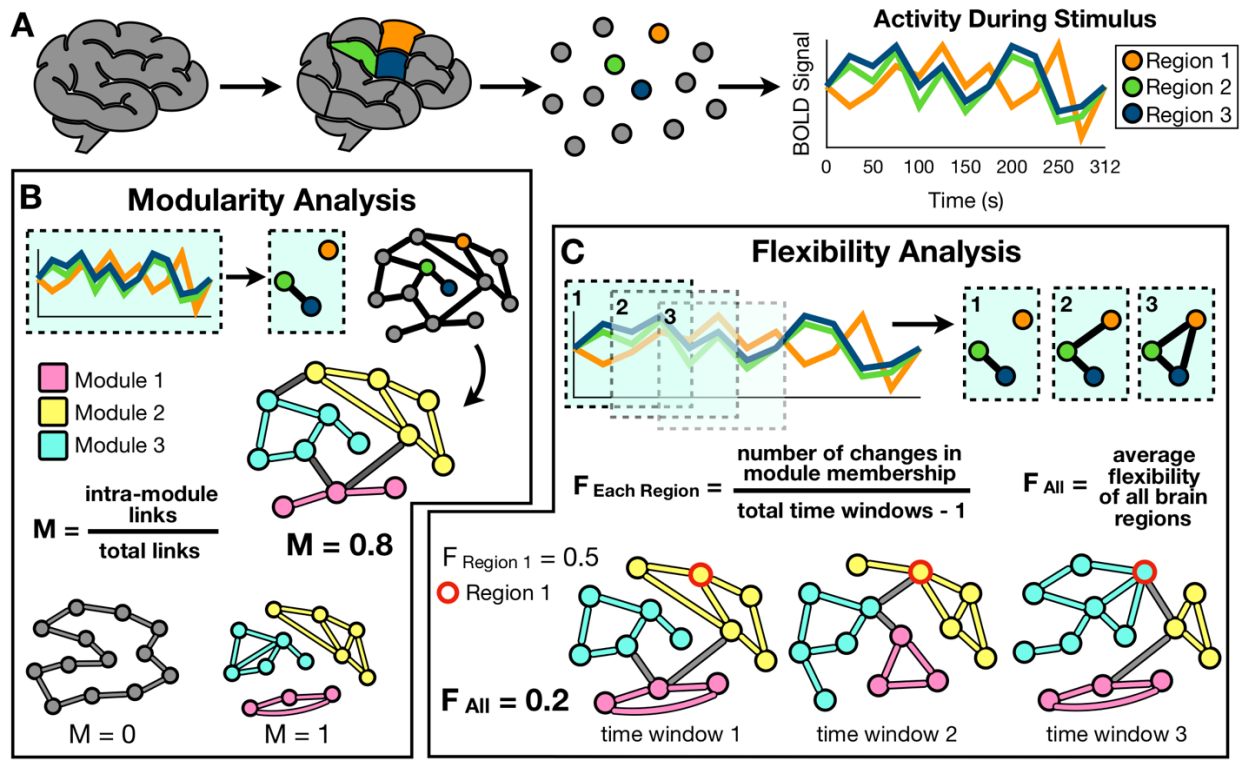

**Figure 1:** The whole-brain network analysis methods. **(A)** The brain is divided into 84 anatomical regions that serve as network nodes. The BOLD signals for each brain region from fMRI are compared to determine the network connections. A link is drawn between two brain regions if their signals are correlated over time: the complete time series is used for modularity, while short overlapping windows of the time series are used for flexibility. **(B) Modularity Analysis:** Modularity is defined as the ratio of links within modules to the total number of links. The main example has three modules with 13 intra-module links of 16 total links, and modularity is therefore 0.8. Example networks for minimum (M=0) and maximum (M=1) modularity are also shown. **(C) Flexibility Analysis:** Flexibility for each brain region is defined as the number of times the brain region changes its module membership, from one time window to the next, divided by the number of time windows. The example network shows that Region 1 changes from the yellow to blue module one time over the two subsequent time windows, and its flexibility is therefore 0.5. The overall flexibility is then the average

of the flexibility values for individual brain regions. More details and full equations are described in the Extended Methods section.

Modularity measures the extent to which the brain regions can be grouped into communities, known as modules, based on sharing many functional connections and having limited connections to the rest of the brain (see **Figure 1A**). In other words, a module contains communities of brain regions that appear to all have highly coordinated activity while processing the stimulus. Modularity is expressed as the number of links inside modules divided by the total links in the network. High modularity means that the network consists of discrete communities that are substantially isolated from each other; in other words, these networks have mostly *intra*-module links. Low modularity means that the communities are less distinct and are substantially connected to other communities; these networks have mostly *inter*-module links.

Meanwhile, flexibility determines how dynamic the network is over the course of the auditory piece based on the rate that each brain region changes its module membership (see **Figure 1B**). To determine flexibility, we used a sliding-window approach and extracted 80 short overlapping portions of the neuroimaging run. A network was constructed for each of these time windows; the modular structure was determined for each network; and flexibility was computed for individual brain regions based on differences in the network modules from one time window to the next. It was calculated as the number of times that a brain region changed its module membership divided by the number of subsequent time windows. High flexibility means that the brain region has a high rate of switching modules; in other words, the brain region is found to be a part of a different community in almost every time window. Low flexibility means that the brain region has a low rate of switching modules; it mostly stays with its same community throughout the entire run. The results were averaged over all brain regions to assign an overall flexibility value to a participant's brain during a particular auditory piece.

The modularity and flexibility measures are based on similar network principles; however, they are not inherently related. In randomly simulated brain networks, the correlation coefficient between the modularity and flexibility plotted for these networks is zero (Ramos-Nuñez et al., 2017). The correlation coefficient ($r$) quantifies whether a straight line can be drawn through the data points, thus determining if there is a significant relationship between the two measures. A strong positive relationship ($r$ close to 1) means that higher modularity is accompanied by higher flexibility, and lower modularity is accompanied by lower flexibility. A stronger negative relationship ($r$ close to -1) means that higher modularity is associated with lower flexibility, and vice versa. A null relationship ($r = 0$) means that there is no overall trend between modularity and flexibility. The modularity-flexibility relationship has not yet been explored during task-based fMRI as we describe here. We were interested in whether this relationship could distinguish differences in how participants processed each auditory piece and, importantly, the culturally familiar music versus the unfamiliar Gagaku.

## 3 Results

When first comparing familiar music and speech, we found distinct modularity-flexibility relationships (see **Figures 2A and 2B**). There were strong negative correlations during Chaplin ($r = -0.68$, p-value = 0.030) and Cronkite ($r = -0.58$, p-value = 0.063). In contrast, the negative correlations were weaker during Self ($r = -0.44$, p-value = 0.032) and Bach ($r = -0.46$, p-value =

0.024). Though we are working on the scale of the whole-brain, this result is consistent with an earlier study of the auditory cortex, which found unique responses for music and speech (Norman-Haignere et al., 2015). This suggests that the modularity-flexibility relationship is able to distinguish different brain states during auditory processing.

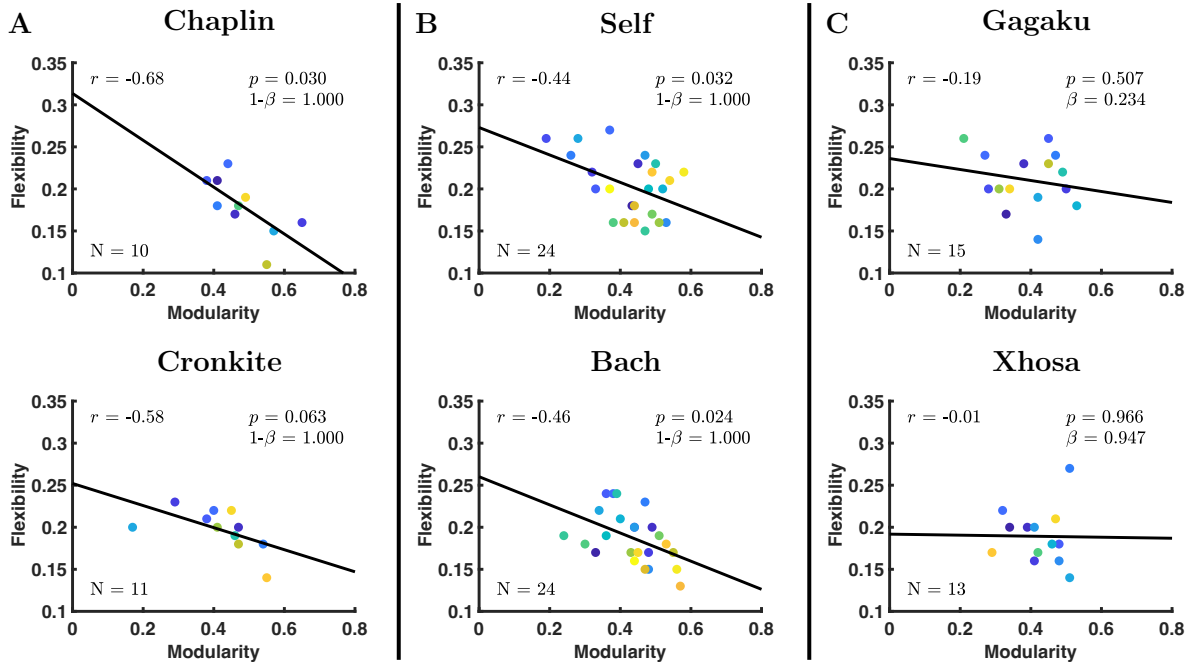

**Figure 2:** The modularity-flexibility relationship during **(A)** English speech (Chaplin and Cronkite), **(B)** self-selected and culturally familiar music (Bach), and **(C)** culturally unfamiliar music (Gagaku) and speech (Xhosa). Data points represent individual participants (colors for each participant are consistent across all six graphs), and $N$ is the number of participants that listened to that piece. Black lines represent linear fits, $r$ is the Pearson correlation coefficient, $p$ indicates the p-value for two-tailed null hypothesis testing of $r = 0$, $1-\beta$ indicates the statistical power to reject the null hypothesis, and $\beta$ is the Type II error rate when the null hypothesis was not rejected.

We then compared the self-selected and Bach with the Gagaku and again found distinct modularity-flexibility relationships (see **Figures 2B and 2C**). The overall negative correlation reported above for Self and Bach was absent during Gagaku: instead, there was no statistically significant trend ($r = -0.19$, p-value = 0.507). A few participants' brains exhibited higher modularity and lower flexibility than the group averages, or vice versa, however, the majority did not, suggesting that the participants' brains were behaving more idiosyncratically than when listening to Self and Bach. The same was true for Xhosa ($r = 0.01$, p-value = 0.966).

We performed power analyses for the Pearson correlation coefficients to determine the probability that our study found statistically significant effects when these effects actually do exist (see Extended Methods section). For Chaplin, Cronkite, Self, and Bach there was very sufficient power ($1-\beta = 1.00$) in the sample sizes for rejecting the null hypothesis, i.e., determining that the correlation coefficients were not $r = 0$. For Gagaku, there was a marginally acceptable Type II error rate ($\beta = 0.234$) for not rejecting the null hypothesis, i.e., correctly concluding that there was no correlation between modularity and flexibility. For Xhosa, however, there was a very high Type II error rate ($\beta = 0.947$) for not rejecting the null hypothesis; in other words, we cannot necessarily conclude there is a zero correlation. However, there is sufficient power to determine that the correlations between modularity and flexibility for Gagaku ($1-\beta = 0.871$) and Xhosa ($1-\beta = 0.996$) are not $r = -0.4$, and we can therefore conclude that the modularity-flexibility relationships are weaker than they are for the other auditory pieces. Additional participants listening to Gagaku and Xhosa would be needed in order to narrow down exactly how weak the correlations are.

Furthermore, we computed the non-parametric Spearman's rank correlation coefficient, $\rho$, for the relationship between each participant's modularity and flexibility during different auditory pieces, and we found similar trends to the Pearson correlations. There is a strong negative correlation that is statistically significant for the two English speech excerpts (Chaplin, $\rho = -0.75$, p-value = 0.013; Cronkite, $\rho = -0.76$, p-value = 0.006). There were weaker negative correlations that was statistically significant for Bach ($\rho = -0.48$, p-value = 0.017) and marginally insignificant for Self ($\rho = -0.32$, p-value = 0.128). The modularity and flexibility correlations were again highly insignificant for Gagaku ($\rho = -0.16$, p-value = 0.566) and Click ($\rho = -0.17$, p-value = 0.582).

In each of these cases, the reported flexibility values were averaged over all brain regions. We then looked at the contributions of individual regions. **Figure 3** shows the flexibilities of each brain region during each auditory piece as averaged over all participants.

Brain regions involved in visual processing and somatosensory function were generally the least flexible across all auditory pieces, indicating that the activity within these respective groups of regions remained synchronized throughout the duration of each auditory piece. The anterior cingulate cortex (BA 33), a core emotion processing region (Pereira et al., 2011), and the parahippocampal gyrus (BA 27) were the most flexible across all auditory pieces, indicating that activity within these regions was not highly correlated with the activity in other brain regions. This suggests that part of the procedure for emotion during an auditory stimulus (as estimated by activity in BAs 27 and 33) is a process independent of the rest of the brain.

There was notably higher flexibility in several regions associated with contentment during all three musical pieces versus during all three speech excerpts. One was the inferior frontal gyrus (BAs 44, 45, 46), which has been implicated in determining musical enjoyment (Koelsch et al., 2006). Additionally, there was higher flexibility in the primary and supplementary motor cortex (BAs 4 and 6, respectively) than during the three speech pieces. The ventral tegmental area, a dopaminergic region where activity is a proxy for pleasure, projects into these brain regions (Hosp et al., 2019). These regions may be more dynamic with which brain regions they interact during music as the participants determine musical enjoyment.

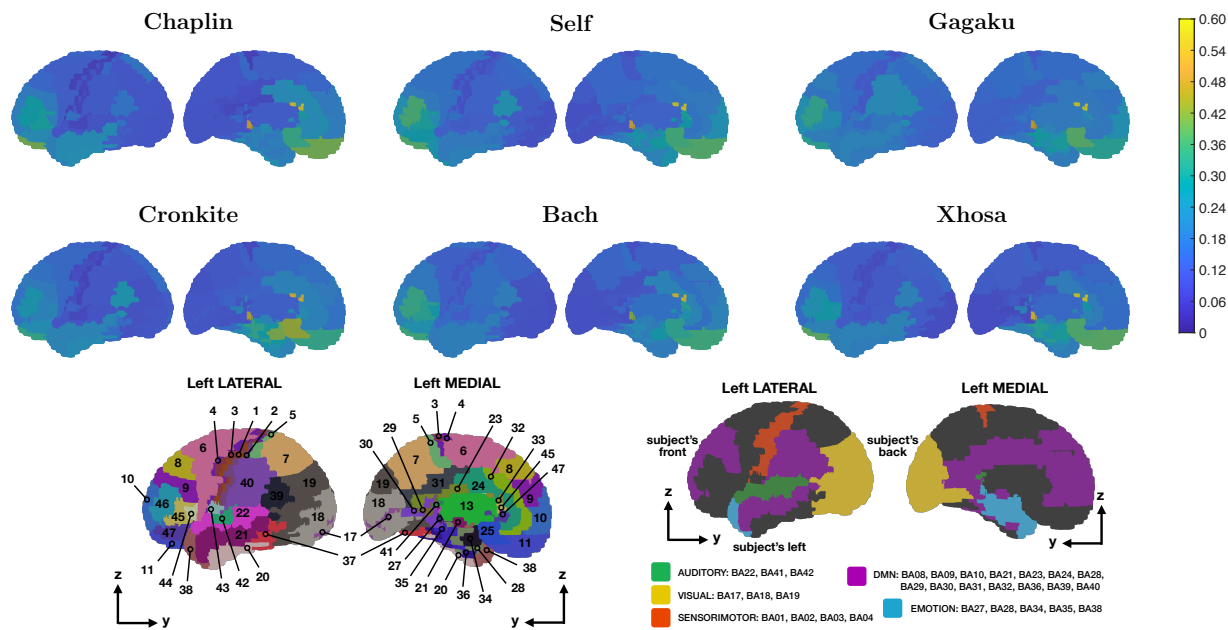

**Figure 3:** The flexibilities for each brain region averaged over all participants. The color bar indicates the flexibility value, which can range from $F = 0$ to 1 ($F = 0.60$ was the maximum in our dataset). $F = 0$ means that the brain region never switched which module it was in throughout the auditory piece; $F = 1$ would mean that the brain region switched its module membership at every time window. The brain figure keys at the bottom show the locations of each BA brain region and several groups of functionally significant BAs.

We took a closer look at the flexibility of the brain regions involved in auditory processing (see **Figure 4**). There was a significant difference in the response of these regions during familiar music and speech, which was in line with our results reported above and those reported by Norman-Haignere et al. (2015). The flexibility was the same in these brain regions during both Self and Bach. Interestingly, there was higher flexibility during the Gagaku than during these familiar pieces. In other words, the auditory regions were more dynamic and had a higher rate of changing modules during Gagaku: the auditory cortex was interacting with more diverse groups of regions across the brain throughout the unfamiliar music in order to process it. It is difficult to assess to what degree these differences were due to the nature of the music itself or to its unfamiliarity—after all, the two are intertwined. Nevertheless, our experiment indicates that whole brain analysis is a promising way to probe the brain's responses to real-world exemplars.

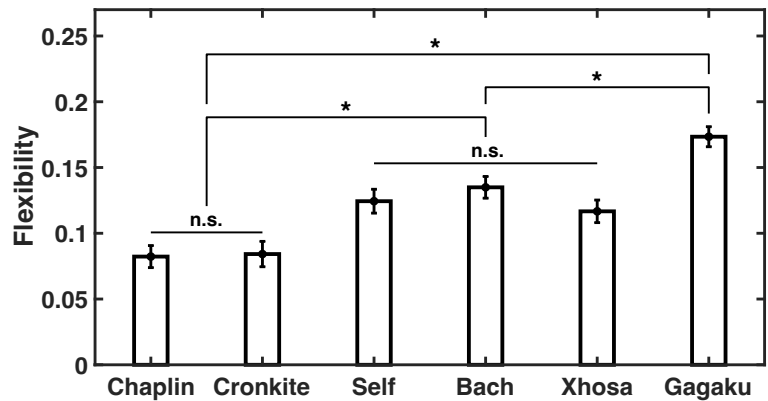

**Figure 4:** The average flexibilities of brain regions involved in auditory processing (BAs 22, 41, 42) for all participants. Error bars indicate standard error. The asterisk indicates a statistical significance of at least p-value < 0.05, and "n.s." indicates that the specified values are not significant. Individual significant p-values are: Gagaku-Self p = 0.001, Gagaku-Bach p = 0.042, Gagaku-Xhosa p < 0.001, Gagaku-Chaplin p < 0.001, Gagaku-Cronkite p < 0.001, Self-Chaplin p = 0.002, Self-Cronkite p = 0.007, Bach-Chaplin p < 0.001, Bach-Cronkite p = 0.001, Chaplin-Xhosa p = 0.006, Cronkite-Xhosa p = 0.017.

## 4 Discussion

Our study analyzed the whole-brain processing of music and speech and quantified the activity using the relationship between modularity and flexibility. We found distinct contrasts in how the brain behaved when listening to speech versus music, as well as Self and Bach versus Gagaku. These results are in line with existing evidence for the impact of enculturation in shaping musical minds (Neuhaus, 2003; Hannon and Trehub, 2005a; Hannon and Trehub, 2005b; Hannon and Trainor, 2007; Morrison et al., 2008; Nan et al., 2008, Morrison and Demorest, 2009; Soley and Hannon, 2010; Cameron et al., 2015; Haumann et al., 2018). To the best of our knowledge, our work is the first to examine differences in the dynamic modular organization of the functional brain network when participants listen to music that differs in cultural familiarity.

During both Self and Bach, there was an overall relationship between modularity and flexibility: the degree of community structure in a participant's brain network was negatively correlated with the dynamics of the communities. However, during Gagaku, there was no significant correlation between these measures. In addition, whereas the flexibility of the auditory cortex was the same during Self and Bach, there was increased flexibility during Gagaku. During an earlier stage of this study, we analyzed brain activation for the first 12 participants (Karmonik et al., 2016). On average over these participants, there was enhanced blood oxygenation level dependent (BOLD) signals, increased functional connectivity among activated voxels, and increased information flow during the three musical excerpts, as compared to the speech pieces. Furthermore, while there was individual variation, the average activation maps for Self, Bach, and Gagaku exhibited distinct traits. For instance, while all three musical examples engaged the primary auditory cortex, Gagaku showed the least overlap with the other stimuli. Self and Bach both activated the superior frontal gyrus, involved in introspective thought (Goldberg et al., 2006), while Gagaku uniquely activated the superior parietal lobule, necessary for working memory (Koenigs et al., 2009). The contrast with Self was particularly noticeable: while, as anticipated, Self activated emotional centers such as the periaqueductal gray and the anterior cingulate cortex (Pereira et al., 2011), the Gagaku did not. Taken as a whole, we found that Self and Bach were more closely related, while Gagaku was an outlier. These previous results, in tandem with our modularity-flexibility analysis here, suggest that the adult brain may treat culturally familiar repertoire in a significantly different way than culturally unfamiliar music.

The modularity-flexibility relationship indicated that processing Self and Bach was more similar to processing the English speech excerpts than it was to processing Gagaku. We hypothesize that this may reflect differences in processing efficiency. For context, previous work found a strong negative correlation between modularity and flexibility ($r = -0.78$) during resting-state fMRI (Ramos-Nuñez et al., 2017), in which participants are not presented with any external stimulus. Since the brain is not being tasked with anything to process, the modularity-flexibility relationship could indicate an optimized, energy-saving network configuration. During a listening task, the brain perceives each auditory stimulus by comparing it to known stimuli templates in its auditory memory (Peretz and Coltheart, 2003; Zatorre and Salimpoor, 2013). Self and Bach, recognized by the brain as familiar music, could promote the negative correlation between modularity and flexibility because the brain can efficiently process these musical pieces. The novel and unpredictable composition of a culturally unfamiliar piece makes it inherently more difficult to process (Zatorre and Salimpoor, 2013). The optimized functional organization may therefore break down during Gagaku, as indicated by the lack of clear modularity-flexibility relationship. The concept can be visualized by imagining different configurations of the brain network representing different possible states. These states can be mapped onto an energy landscape, which is graph of the energy associated with all possible configurations of a complex system. In our case, each configuration state is characterized by a modularity value, which represents the depth of the state, and a flexibility value, which represents the rate that the brain transitions from one state to another. Given that higher modularity makes a complex system more stable (Simon, 1962) and more energy is needed to overcome a larger barrier between states, the system is more likely to transition from shallow states (low modularity, high flexibility) and less likely to transition out of deeper states (high modularity, low flexibility). This qualitative perspective was first put forward by Ramos-Nuñez et al. (2017). Applying this perspective to our work, a negative modularity-flexibility relationship is assumed to be an optimal configuration for the brain to function efficiently and minimize energy. When this relationship is not present, the brain is either using extra energy to try to quickly explore the landscape and transition between deep states (high modularity, high flexibility), or it is inefficiently exploring the landscape by transitioning slowly between shallow states (low modularity, low flexibility). A more rigorous theoretical investigation of this hypothesis is needed and could be accomplished utilizing the mathematical formalism commonly employed to study the energy landscape of complex systems in condensed matter (Pietrucci, 2017).

Researchers have speculated about the degree to which the neural communities involved in music perception are task-specific or shared with other cognitive tasks. For instance, music and language processing show considerable overlap, especially during early childhood (Patel, 2012; Patel, 2015; Patel and Morgan, 2017; Brandt et al., 2012). Likewise, while musical vernaculars differ widely between cultures, Mehr et al. (2019) have shown that naïve listeners often successfully rely on affective cues such as complexity and levels of arousal to accurately interpret the function of unfamiliar songs—succeeding precisely because these affective cues are related to social cues in everyday life. For the adult brain, listening to familiar repertoire has likely been streamlined into more domain-specific networks, which may explain the equal flexibility of the auditory cortex that we observed during Self and Bach. The higher flexibility of the auditory cortex that we found during Gagaku may suggest that participants' brains are utilizing alternative cues in order to decipher the culturally unfamiliar music.

It is worth noting that, the Bach and Gagaku have many other differences in addition to the familiarity of their musical languages—for instance, the Bach is a piano piece, but the Gagaku is written for an ensemble of Asian instruments. While previous studies have mapped the subsections of the auditory cortex that process specific musical features (Woods et al., 2009; Leaver and Rauschecker, 2010; Norman-Haignere et al., 2015), our study analyzed brain activity at a larger resolution in which the auditory cortex was treated as a homogenous brain region. We hypothesize that the whole-brain network approach would not be sensitive enough to distinguish differences in traditional acoustical properties, such as pitch and timbre. In any case, consideration of the differences in acoustical properties of the stimuli that impact brain activation at the cellular level is less relevant to the aim of present study: to study how the brain, as a complex system, adapts its functional organization while listening to a well-known song (Self), an unknown song that contains musical features customary to the listener (Bach), and an unknown song that contains musical features that are uncustomary to the listener (Gagaku)." It is often difficult to interpret how activations in particular regions integrate with the activity of the rest of the brain to holistically process a stimulus (de-Wit et al., 2016), and so by taking a whole-brain network approach, we are able to quantify differences in the large-scale architecture of different brain states during music listening. Additionally, isolated instances of pitch or other acoustic features cannot necessarily be culturally grounded (Morrison and Demorest, 2009), and so by presenting longer stimuli, our participants have a greater awareness of the musical character and continuity of each selection.

## 5    Limitations

There are a few limitations of the present work. First, this is a small-scale, pilot study. The results we present offer a promising new analysis approach for the field and would be strengthened by including more participants, furthermore, to adequately address the potential influence of age and gender. Second, given the scale of the study, we were limited in the number of stimuli we could test. Earlier studies found that even when testing culturally different repertoires on participants of different nationalities, a lack of difference in activation patterns can result when the musical features of the repertoires are to similar (Demorest and Morrison, 2003). We therefore aimed for strongly contrasting ones, feeling that those would provoke the most measurable results. However, this introduces confounds and makes it hard to ascribe the brain behavior to any single cause. Our methodology in follow-up work will be refined to create more overlap in surface musical features, such as instrumentation, to further tease out the effects of familiarity and unfamiliarity. Third, we were certain that Bach was representative of recognizable classical music and that Gagaku was representative of an unknown repertoire to the participants in this study. However, the collection of psychological measures to specifically quantify music familiarity, comprehension, preference, and emotional stimulation would allow us to further probe how conscious experience contributes to differences in whole-brain processing. This would also open the door for distinguishing how individual participants process the various auditory pieces. Finally, an additional caveat of our study is that many of the self-selected tracks chosen by our participants contained lyrics. In future studies, it would be advantageous to probe the impact of mixing language and music.

# 6 Future Research

Scientists have long speculated the degree to which music cognition is innate or acquired through exposure and learning. Our study helps lay the groundwork for further research into this question. An increasing number of cross-cultural music studies are being conducted, and while a few involve neuroimaging (for instance, Demorest and Morrison, 2003; Nan et al., 2008; Demorest, 2009), more are needed to explore the varied brain responses during music cognition. Some researchers note that it is increasingly difficult to find listeners who have never heard Western music (McDermott et al., 2016; Stevens 2012). However, cross-cultural studies need not include Western music as one of the musical traditions being analyzed (Jacoby et al., 2020). Music familiarity and brain response from any two, or more, cultures could be compared. In fact, investigations into the extent to which cultural exposure shapes music cognition in a wide diversity of populations will be crucial contributions to this field of research.

Seeing as our study was limited to those accustomed to Western music, a logical next step would be to perform the same study with aficionados of Gagaku music, as well as those conversant with both Gagaku and Western musical traditions. It would be interesting to see how the brain responses of each of these groups compared. Wong et al. (2009) performed memory and recognition tasks with participants raised with exposure to both Indian and Western musical traditions and compared them with participants familiar with one musical culture or the other. While most participants showed an in-culture bias, those who were familiar with both Indian and Western music exhibited equal brain responses to the music of both cultures (Wong et al., 2009). We would be interested to see how the activation patterns and whole-brain network measures of experts of Gagaku compare to those who were completely unaccustomed to it.

Our study deliberately excluded trained musicians, but it would be salient to compare them to untrained listeners. Demorest and Morrison (2003) found increased activation in both Chinese and American performers in response to culturally familiar and unfamiliar stimuli as compared to untrained listeners. It would be of particular interest to explore whether the negative correlation between modularity and flexibility also breaks down for professional musicians when listening to culturally unfamiliar music.

# 7 Conclusion

In summary, we studied whole-brain activity from fMRI of a group of healthy adult participants while they listened to various music and speech pieces. Our complex systems theory approach and use of longer excerpts of real-world stimuli allowed us to explore differences in music processing of culturally familiar versus culturally unfamiliar music. While there were significant trends in network modularity and flexibility during Self and Bach, there was no trend during Gagaku. There was also a noteworthy increase in the flexibility of the auditory cortex for all participants during Gagaku, which suggests that participants' brains were drawing upon novel resources to decipher this music.

Studying the whole brain enables us to study the complex synergies between different brain regions and examine the degree to which it is the same for everyone. This is germane in areas such as music therapy, which must concern itself with the degree to which musical interventions can be generalized or must be customized for individual patients. Indeed, our results suggest that music processing may take individual components of musicality and assemble them into interacting communities based on both cultural exposure and personal preferences.

Because it is culturally omnipresent yet enormously varied, music offers a particularly revealing window into how our brain engages with experiences both familiar and new. Our work demonstrates the utility of the modularity and flexibility measures of whole-brain network activity to quantify the complex neural operations occurring during music perception and to propose a theoretical grounding for why the brain organizes and reorganizes itself during different types of music. As Eagleman (2020) writes, "For humans at birth, the brain is remarkably unfinished, and interaction with the world is necessary to complete it" (p. 20). By using real world samples and whole brain analysis, we can better understand how those interactions with the world shape our musical brains.

## 8 Extended Methods

### 8.1 Musical Feature Extraction

To quantify the difference in human perception of these auditory pieces, as a proxy for differences in the overall listening experience, we utilized the Rhythm Pattern feature extractor (Lidy and Schindler, 2016; Lidy and Rauber, 2005; Rauber et al., 2003). This extractor calculates the similarity between auditory signals by quantifying the combined human perception of rhythm, pitch/melody, and timbre information during an auditory piece. The feature set was developed to capture differences in psychoacoustic phenomena while listening to music, which we feel is more meaningful to our present study than differences in acoustic statistics calculated directly from the auditory waveform.

Briefly, a Fourier transform is first computed for the audio signal, and the frequencies are grouped into 24 psycho-acoustically motivated critical-bands on the perceptual Bark scale (Zwicker 1961). Spectral masking is then performed to reproduce the phenomena of quiet sounds being occluded from human hearing by louder sounds that are present simultaneously, closely before, or closely after. Further processing is performed, including transformation into different perceived loudness scales. Another discrete Fourier Transform is computed to create a time-invariant representation of the spectrum, known as the Rhythm Pattern. This quantifies the amplitude modulations of the loudness in individual Bark scale bands. These modulations of the loudness occur at different frequencies; the algorithm cuts off at 10 Hz, which corresponds to a modulation of 600 bpm in the Bark scale bands. The amplitude modulations are binned into 60 modulation frequencies per each of the 24 Bark bands, leading to a 1440-dimensional feature vector. The Rhythm Pattern results for Bach and Gagaku are shown in **Figure 5**.

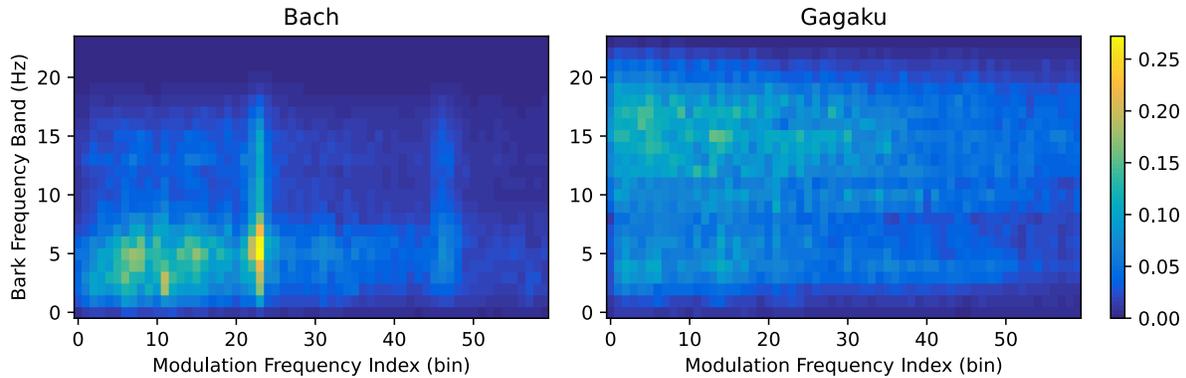

**Figure 5:** Rhythm Patterns calculated using the algorithm developed by Lidy and Schindler (2016) (Lidy and Rauber, 2005; Rauber et al., 2003) for Bach and Gagaku. The Bark frequencies are 24 bands determined from psychoacoustic testing to be important to human perception (Zwicker 1961). The colormap represents the amplitude modulation at the binned frequency on the x-axis (bin 0 is a modulation frequency of 0 Hz and bin 60 is a modulation of 10 Hz).

Given the high dimensionality of these data (Aggarwal et al., 2001), the Manhattan distance was used to calculate the distance between musical feature vectors. A larger Manhattan distance means that the auditory pieces were more musically distinct. The average distance from Bach to Self songs was $869 \pm 31$ (standard error of mean). The average distance from Gagaku to Self songs was $1007 \pm 38$, and the distance from Gagaku to Bach was 1302.

## 8.2   Neuroimaging

Data were acquired from 25 participants (ages 18 to 82, nine males) recruited from the Houston community that were not taking chronic medication or psychoactive drugs. The group was heterogeneous in gender, age, and degree of music education. The Houston Methodist Hospital Institutional Review Board approved the research, and informed consent was obtained from all participants. Imaging data were collected at the MRI core of Houston Methodist Research Institute on a Philips Ingenia 3.0T scanner. Scans for the first 12 participants were acquired during an earlier stage of the study (Karmonik et al., 2016; Karmonik et al., 2020). Anatomical reference scans used a turbo field echo pulse sequence with a field of view of 24x24x16.5cm (1.0mm isotropic resolution, 8.2ms repetition time, 3.8ms echo time). The task-based functional scans used an echo planar imaging pulse sequence with a field of view of 22x22x12cm for 130 brain volumes in each run (1.5x1.5x3.0mm resolution, axial orientation, 2400ms repetition time, 35ms echo time).

Based on the fMRI repetition time, each volume represented 2.4s of activity. The listening task followed a block design in which each scan began with 24s of silence, followed by 12 blocks of alternating auditory stimulus and silence (each for 24s), for a total run of 312s (130 volumes). The first 24s of silence at the start of each run were not used in analyses, leaving a 288s time series with 120 volumes. Analyses were conducted across the 288s time series without specific regard for individual blocks. We were interested in looking at the network architecture of the sustained brain state as a whole during each run. It is assumed that the brain is processing both auditory features and psychological response (e.g., emotion, pleasure, memory) when the stimulus is on, and only the latter when the stimulus is off. The block design encourages better cognitive engagement, whereas presentation of the uninterrupted auditory stimulus for 2min 24s (144s) may lead to less active listening during portions of the time series that would differ among participants. Additionally, previous work with this dataset was able to determine stimulus-specific activation in regions of interest using the silence block as a control condition (Karmonik et al., 2016). The number of auditory pieces that each participant listened to varied depending on the participant's tolerance for the total MRI scan duration. Data from one participant were excluded due to technical difficulty. In the analyzed data, 24 participants listened to Self, 24 listened to Bach, 15 listened to Gagaku, 13 listened to Xhosa, 11 listened to Cronkite, and 10 listened to Chaplin.

Standard preprocessing steps were taken to reduce artifacts in the fMRI data, including correcting for motion, constant offset, and high-frequency contributions in the BOLD signal, and implemented as described in (Karmonik et al., 2016). To construct the whole-brain network, functional and anatomical MRI scans were combined using AFNI software (Cox, 2012) and transformed into Talairach coordinates, which spatially warps each participant's brain image to a standardized, three-dimensional space. The brain was then parcellated into 84 Brodmann areas (BAs). In prior work, the functional network was constructed using other parcellation schemes and consistent trends were observed in modularity analyses (Yue et al., 2017). Functional connectivity between BAs was determined by calculating pairwise Pearson correlation coefficients of BOLD time series. The undirected functional connectivity matrix was then binarized to keep the top 400 edges, 11.5% network density, to improve the signal-to-noise ratio (Yue et al., 2017; Chen and Deem, 2015).

### 8.3  Quantifying Brain Response

Modularity, $M$, is calculated from the inferred functional networks of each participant as they listened to each auditory piece using Newman's algorithm (Newman, 2006; Chen and Deem, 2015). The algorithm assigns BAs into modules, $\sigma$, based on the configuration that maximizes $M$ defined as

$$M(\{\sigma\}) = \frac{1}{2L} \sum_k \sum_{ij \in \sigma_k} \left( A_{ij} - \frac{a_i a_j}{2L} \right)$$

where $L$ is the total number of links in the network, $A_{ij}$ is 1 if there is a link between BA $i$ and BA $j$ or otherwise 0, and $a_i$ is the total number of links of BA $i$. The inner sum is evaluated for all $ij$ pairs of BAs in module $k$, $\sigma_k$, and the outer sum is evaluated for all modules in the network. The $a_i a_j /2L$ term reduces $M$ by the modularity that would be expected in a random network.

The number and composition of modules are not predefined, but rather deduced by the algorithm based on the grouping of brain regions that optimizes the function for $M$. Briefly, the functional connectivity matrix is input into the algorithm, and the algorithm takes a top-down approach in determining modules. The nodes are first divided into two groups based on the largest eigenvector of the connectivity matrix, and modularity is calculated from this arrangement. Next the algorithm checks what the modularity would be by moving each node between groups. If modularity increases (above a threshold of 0.01), the best move of nodes that maximizes modularity is performed. The algorithm then attempts to subdivide the groups and repeats testing how modularity would change when each node is moved among groups. This continues until subdividing the groups further does not increase the calculated modularity. The resulting groups are determined to be the functional modules. This data-driven approach means that there is no bias in assigning brain regions into particular modules based on any assumed functional relationships. Since the modules are not universally established or constrained, this allows us to meaningfully compute the modularity and the rate that the modules are reorganized for different auditory pieces.

Flexibility, *F*, is defined by the average rate that BA nodes changed their module allegiance, as determined using a sliding window approach (Bassett et al., 2011; Hutchinson et al., 2013). Consistent with previous studies, a 40-volume window is used, which is approximately where the time series autocorrelations return to zero (Ramos-Nuñez et al., 2017). The windows are moved forward one volume at a time until reaching the end of the 120-volume time series. A network is constructed in each window, and Newman's algorithm is used to determine the assignment of BAs to modules. Since the algorithm may artificially label the same module differently in two subsequent windows, the relabeling process devised by Ramos-Nuñez et al. (2017) is used. The average flexibility *F* for all brain regions is then calculated as,

$$F = \frac{1}{N(T-1)} \sum_i^N \sum_t^T (1 - \delta_{m_{i,t} m_{i,t+1}})$$

where *N* = 84 is the number of BAs, *T* = 80 is the number of time windows of 40 volumes each, and $m_{i,t}$ is the module assignment of BA *i* in window *t*. The $\delta$ is the Kronecker delta, which is 1 if the module assignment of BA *i* is the same in time windows *t* and *t+1* or otherwise 0. The *T-1* is included in the denominator to scale flexibility between 0 and 1, where *F=1* means that all brain regions switched which module that they were in at every consecutive time window, and *F=0* means that the division of brain regions into modules is completely static over the duration of the stimulus. When analyzing the flexibility of individual BAs, *F* is calculated without computing the average over *N*.

We were motivated to look at the modularity-flexibility relationship to study how the brain processes different auditory pieces because the degree of dynamic, modular structure in brain networks is associated with differences in cognitive performance under different task demands. Prior theory modeled the benefit of high modularity for performing fast, simple cognitive tasks and the benefit of low modularity for longer, more complex tasks (Chen and Deem, 2015), and experiments have demonstrated this dichotomous connection between performance and both resting-state (Yue et al., 2017) and task-based (Lebedev et al., 2018) modularity. The opposite relationship has been experimentally observed for flexibility, where low flexibility correlates with performance on simple tasks, and high flexibility correlates with performance on complex tasks (Ramos-Nuñez et al., 2017). Furthermore, there is a negative relationship between modularity and flexibility in resting-state fMRI data (Ramos-Nuñez et al., 2017). Ramos-Nuñez et al. (2017) previously put forward a dynamical systems perspective, in which different organizations of the functional network represent different attractor states, to intuitively explain this relationship: the modularity of the network represents the depth of the state, and flexibility represents the rate that the brain transitions from one attractor state to another; the system is less likely to transition out of deeper states and more likely to transition from shallow states. Using this perspective, a negative modularity-flexibility relationship is assumed to be an optimal configuration for the brain to minimize energy and efficiently process stimuli.

### 8.4 Statistical Analyses

All statistics and hypothesis testing were carried out using the functions available in the MATLAB Statistics and Machine Learning Toolbox (MATLAB, 2020). In our interpretation of the results, we considered a p-value < 0.10 as an acceptable Type I error rate for rejecting the null hypothesis, where p-value < 0.05 is considered statistically significant and p-value < 0.10 is marginally significant. We considered $1-\beta > 0.8$ as sufficient statistical power and $\beta < 0.2$ as an acceptable Type II error rate for not rejecting the null hypothesis.

In particular for the power analyses of the modularity-flexibility relationship effect sizes, we calculated the power achieved for the number of participants who listened to each auditory piece, when the true values are the observed Pearson correlation coefficients ($r$), and the null hypothesis is $r = 0$. For the auditory pieces in which we did not reject the null hypothesis (Gagaku and Xhosa), we calculated the Type II error rate $\beta$ as 1 – power. We conducted brief secondary power analyses on Gagaku and Xhosa to assess if the modularity-flexibility relationship effect sizes for these pieces ($r = -0.19$ and $r = -0.01$, respectively) were weaker than those of the other auditory pieces (all having $r < -0.4$); the statistical power achieved when the null hypothesis being rejected is $r = -0.4$ was calculated.

## 9 Conflict of Interest

The authors declare that the research was conducted in the absence of any commercial or financial relationships that could be construed as a potential conflict of interest.

## 10 Compliance with ethical standards

The authors declare that all experiments on human subjects were conducted in accordance with the Declaration of Helsinki and that all procedures were carried out with the adequate understanding and written consent of the subjects. The authors also certify that formal approval to conduct the experiments described has been obtained from the human subjects review board of their institution and could be provided upon request.

## 11 Author Contributions

A.K.B., J.T.F., and C.K. designed the fMRI experiments. A.K.B. and J.T.F. selected and prepared the auditory stimuli. C.K. carried out fMRI collection and data preprocessing. M.E.B. developed the theoretical framework and analyzed the data. A.K.B. and M.E.B. interpreted the results and wrote the manuscript. All authors discussed the results and provided critical feedback on the final manuscript.

## 12 Funding

The research was supported by the Center for Theoretical Biological Physics at Rice University (National Science Foundation, PHY 1427654), the Ting Tsung and Wei Fong Chao Foundation, and the Houston Methodist Center for Performing Arts Medicine. M.E.B. additionally supported by a training fellowship from the Gulf Coast Consortia, on the NLM Training Program in Biomedical Informatics & Data Science (T15LM007093).

## 13 Acknowledgments

The authors would like to thank Fengdan Ye for helpful discussions about the theory of this work. Part of this work appears in M.E.B.'s doctoral thesis (Bonomo, 2020).

## 14 Data Availability Statement

The datasets generated and analyzed during the current study are available from the corresponding author on reasonable request.

## 15 References

Aggarwal C.C., Hinneburg A., and Keim D.A. (2001). On the Surprising Behavior of Distance Metrics in High Dimensional Space. In: Van den Bussche J., Vianu V. (eds) Database Theory — ICDT 2001. ICDT 2001. Lecture Notes in Computer Science, vol 1973. Springer, Berlin, Heidelberg. https://doi.org/10.1007/3-540-44503-X_27

Bassett, D. S., Wymbs, N. F., Porter, M. A., Mucha, P. J., Carlson, J. M., and Grafton, S. T. (2011). Dynamic reconfiguration of human brain networks during learning. Proceedings of the National Academy of Sciences USA. 108, 7641–7646. https://doi.org/10.1073/pnas.1018985108

Brandt, A., Gebrian, M., and Slevc, L. R. (2012). Music and early language acquisition. Frontiers in Psychology, 3(SEP). https://doi.org/10.3389/fpsyg.2012.00327

Bonomo, M. E. (2020). Investigating Modular Structure and Function in Biology: from Immunology to Cognition. Doctoral dissertation, Rice University. https://hdl.handle.net/1911/109615

Bonomo, M. E., Karmonik, C., Brandt, A. K., and Frazier, J. T. (2020). Modularity allows classification of human brain networks during music and speech perception. Preprint. arXiv:2009.10308. https://arxiv.org/abs/2009.10308

Cameron, D. J., Bentley, J., and Grahn, J. A. (2015). Cross-cultural influences on rhythm processing: Reproduction, discrimination, and beat tapping. Frontiers in Psychology, 6(MAR). https://doi.org/10.3389/fpsyg.2015.00366

Chen, M., and Deem, M. W. (2015). Development of modularity in the neural activity of children's brains. Physical Biology. 12, 016009. https://doi.org/10.1088/1478-3975/12/1/016009

Cox, R. W. (2012). AFNI: what a long strange trip it's been. Neuroimage. 62, 743–747. https://doi.org/10.1016/j.neuroimage.2011.08.056

de-Wit, L., Alexander, D., Ekroll, V., and Wagemans, J. (2016). Is neuroimaging measuring information in the brain? Psychonomic Bulletin & Review. 23, 1415–1428. https://doi.org/10.3758/s13423-016-1002-0

Demorest, S. M., and Morrison, S. J. (2003). Exploring the Influence of Cultural Familiarity and Expertise on Neurological Responses to Music. In Annals of the New York Academy of Sciences (Vol. 999, pp. 112–117). New York Academy of Sciences. https://doi.org/10.1196/annals.1284.011

Demorest, S. M., Morrison, S. J., Stambaugh, L. A., Beken, M., Richards, T. L., and Johnson, C. (2009). An fMRI investigation of the cultural specificity of music memory. Social Cognitive and Affective Neuroscience, 5(2–3), 282–291. https://doi.org/10.1093/scan/nsp048

Eagleman, D. (2020). Livewired: The Inside Story of the Ever-changing Brain. New York: Pantheon.

Goldberg, I. I., Harel, M., and Malach, R. (2006). When the brain loses its self: prefrontal inactivation during sensorimotor processing. Neuron. 50, 329–339. https://doi.org/10.1016/j.neuron.2006.03.015

Hannon, E. E., and Trehub, S. E. (2005a). Metrical categories in infancy and adulthood. Psychological Science, 16(1), 48–55. https://doi.org/10.1111/j.0956-7976.2005.00779.x

Hannon, E. E., and Trehub, S. E. (2005b). Tuning in to musical rhythms: Infants learn more readily than adults. Proceedings of the National Academy of Sciences of the United States of America, 102(35), 12639–12643. https://doi.org/10.1073/pnas.0504254102

Hannon, E. E., and Trainor, L. J. (2007, November). Music acquisition: effects of enculturation and formal training on development. Trends in Cognitive Sciences. https://doi.org/10.1016/j.tics.2007.08.008

Haumann, N. T., Kliuchko, M., Vuust, P., and Brattico, E. (2018). Applying acoustical and musicological analysis to detect brain responses to realistic music: A case study. Applied sciences, 8(5), 716. https://doi.org/10.3390/app8050716

Honing, H., ten Cate, C., Peretz, I. and Trehub, S.E. (2015). Without it no music: cognition, biology and evolution of musicality. Philosophical Transactions of the Royal Society B: Biological Sciences, 370(1664), 20140088. https://doi.org/10.1098/rstb.2014.0088

Hosp, J.A., Coenen, V.A., Rijntjes, M., Egger, K., Urbach, H., Weiller, C. and Reisert, M. (2019). Ventral tegmental area connections to motor and sensory cortical fields in humans. Brain Structure and Function. 224(8), 2839-2855. https://doi.org/10.1007/s00429-019-01939-0

Hutchison, R. M., Womelsdorf, T., Allen, E. A., Bandettini, P. A., Calhoun, V. D., Corbetta, M., Della Penna, S., Duyn, J. H., Glover, G. H., Gonzalez-Castillo, J., Handwerker, D. A., Keilholz, S., Kiviniemi, V., Leopold, D. A., de Pasquale, F., Sporns, O., Walter, M., and Chang, C. (2013). Dynamic functional connectivity: promise, issues, and interpretations. Neuroimage. 80, 360–378. https://doi.org/10.1016/j.neuroimage.2013.05.079

Jacoby, N., Margulis, E. H., Clayton, M., Hannon, E., Honing, H., Iversen, J., Klein, T. R., Mehr, S. A., Pearson, L., Peretz, I., Perlman, M., Rainer, P., Ravignani, A., Savage, P. E., Steingo, G., Stevens, C. J., Trainor, L., Trehub, S., Veal, M., and Wald-Fuhrmann, M. (2020). Cross-cultural work in music cognition: Challenges, insights, and recommendations. Music Perception, 37(3), 185–195. https://doi.org/10.1525/MP.2020.37.3.185

Karmonik, C., Brandt, A., Anderson, J. R., Brooks, F., Lytle, J., Silverman, E., and Frazier, J. T. (2016). Music Listening Modulates Functional Connectivity and Information Flow in the Human Brain. Brain Connectivity, 6(8), 632–641. https://doi.org/10.1089/brain.2016.0428

Karmonik, C., Brandt, A., Elias, S., Townsend, J., Silverman, E., Shi, Z., and Frazier, J. T. (2020). Similarity of individual functional brain connectivity patterns formed by music listening quantified with a data-driven approach. International Journal of Computer Assisted Radiology and Surgery, 15(4), 703–713. https://doi.org/10.1007/s11548-019-02077-y

Koelsch, S., Fritz, T., v. Cramon, D. Y., Müller, K., and Friederici, A. D. (2006). Investigating emotion with music: an fMRI study. Human Brain Mapping. 27, 239–250. https://doi.org/10.1002/hbm.20180

Koenigs, M., Barbey, A. K., Postle, B. R., and Grafman, J. (2009). Superior parietal cortex is critical for the manipulation of information in working memory. Journal of Neuroscience. 29, 14980–14986. https://doi.org/10.1523/jneurosci.3706-09.2009

Leaver, A.M. and Rauschecker, J.P. (2010). Cortical Representation of Natural Complex Sounds: Effects of Acoustic Features and Auditory Object Category. Journal of Neuroscience. 30 (22) 7604-7612. https://doi.org/10.1523/JNEUROSCI.0296-10.2010

Lebedev, A. V., Nilsson, J., and Lövdén, M. (2018). Working memory and reasoning benefit from different modes of large-scale brain dynamics in healthy older adults. Journal of Cognitive Neuroscience. 30, 1033–1046. https://doi.org/10.1162/jocn_a_01260

Lidy T. and Rauber A. (2005). Evaluation of feature extractors and psycho-acoustic transformations for music genre classification. In Proceedings of the Sixth International Conference on Music Information Retrieval. 34-41.

Lidy T. and Schindler A. (2016). Rhythm Pattern music feature extractor by IFS @ TU-Vienna. GitHub repository. https://github.com/tuwien-musicir/rp_extract

MATLAB. (2020). version 9.9.0.1495850 (R2020b). Natick, Massachusetts: The MathWorks Inc.

McDermott, J. H., Schultz, A. F., Undurraga, E. A., and Godoy, R. A. (2016). Indifference to dissonance in native Amazonians reveals cultural variation in music perception. Nature, 535(7613), 547–550. https://doi.org/10.1038/nature18635

Mehr, S. A., Singh, M., Knox, D., Ketter, D. M., Pickens-Jones, D., Atwood, S., Lucas C., Jacoby, N., Egner, A. A., Hopkins, E. J., Howard, R. M., Hartshorne, J. K., Jennings, M. V., Simson, J., Bainbridge, C. M., Pinker, S., O'Donnell, T. J., Krasnow, M. M., and Glowacki, L. (2019). Universality and diversity in human song. Science, 366(6468). https://doi.org/10.1126/science.aax0868

Morrison, S. J., Demorest, S. M., and Stambaugh, L. A. (2008). Enculturation effects in music cognition: The role of age and music complexity. Journal of Research in Music Education, 56(2), 118–129. https://doi.org/10.1177/0022429408322854

Morrison, S. J., and Demorest, S. M. (2009). Cultural constraints on music perception and cognition. Progress in Brain Research. https://doi.org/10.1016/S0079-6123(09)17805-6

Nan, Y., Knösche, T. R., Zysset, S., and Friedend, A. D. (2008). Cross-cultural music phrase processing: An fMRI study. Human Brain Mapping, 29(3), 312–328. https://doi.org/10.1002/hbm.20390

Neuhaus, C. (2003). Perceiving Musical Scale Structures. Annals of the New York Academy of Sciences, 999(1), 184–188. https://doi.org/10.1196/annals.1284.026

Newman, M. E. (2006). Modularity and community structure in networks. Proceedings of the National Academy of Sciences USA. 103, 8577–8582. https://doi.org/10.1073/pnas.0601602103

Norman-Haignere, S., Kanwisher, N. G., and McDermott, J. H. (2015). Distinct Cortical Pathways for Music and Speech Revealed by Hypothesis-Free Voxel Decomposition. Neuron, 88(6), 1281–1296. https://doi.org/10.1016/j.neuron.2015.11.035

Patel, A. D. (2012). Music, Language, and the Brain. Music, Language, and the Brain (pp. 1–526). Oxford University Press. https://doi.org/10.1093/acprof:oso/9780195123753.001.0001

Patel, A. D. (2015). Sharing and Nonsharing of Brain Resources for Language and Music. In Language, Music, and the Brain (pp. 329–356). The MIT Press. https://doi.org/10.7551/mitpress/9780262018104.003.0014

Patel, A. D., and Morgan, E. (2017). Exploring Cognitive Relations Between Prediction in Language and Music. Cognitive Science, 41, 303–320. https://doi.org/10.1111/cogs.12411

Patel, A.D. (2019). Evolutionary Music Cognition: Cross-Species Studies. In *Foundations of Music Psychology: Theory and Research*, ed. P. J. Rentfrow and D. J. Levitin (Cambridge: MIT Press), pp 459-502.

Pereira, C. S., Teixeira, J., Figueiredo, P., Xavier, J., Castro, S. L., and Brattico, E. (2011). Music and emotions in the brain: familiarity matters. PLoS one. 6, e27241. https://doi.org/10.1371/journal.pone.0027241

Peretz, I., and Coltheart, M. (2003). Modularity of music processing. Nature Neuroscience. 6, 688–691. https://doi.org/10.1038/nn1083

Pietrucci, F. (2017). Strategies for the exploration of free energy landscapes: Unity in diversity and challenges ahead. Reviews in Physics. 2, 32-45. https://doi.org/10.1016/j.revip.2017.05.001

Ramos-Nuñez, A. I., Fischer-Baum, S., Martin, R. C., Yue, Q., Ye, F., and Deem, M. W. (2017). Static and Dynamic Measures of Human Brain Connectivity Predict Complementary Aspects of Human Cognitive Performance. Frontiers in Human Neuroscience. 11, 420. https://doi.org/10.3389/fnhum.2017.00420

Rauber, A., Pampalk, E. and Merkl, D. (2003). The SOM-enhanced JukeBox: Organization and visualization of music collections based on perceptual models. Journal of New Music Research. 32(2), 193-210. https://doi.org/10.1076/jnmr.32.2.193.16745

Simon, H. A. (1962). The architecture of complexity. Proceedings of the American Philosophical Society, 106(6), 467–482. https://www.jstor.org/stable/985254

Soley, G., and Hannon, E. E. (2010). Infants Prefer the Musical Meter of Their Own Culture: A Cross-Cultural Comparison. Developmental Psychology. 46(1), 286–292. https://doi.org/10.1037/a0017555

Sporns, O., and Betzel, R. F. (2016). Modular brain networks. Annual Review of Psychology. 67, 613–640.

Stevens, C. J. (2012, October). Music Perception and Cognition: A Review of Recent Cross-Cultural Research. Topics in Cognitive Science. https://doi.org/10.1111/j.1756-8765.2012.01215.x

Tamba, A. (1976). Aesthetics in the Traditional Music of Japan. *The World of Music,* Vol. 18, No. 2, pp. 3-10.

Tanaka, K. and Koto, T. (2016). Traditional Japanese Music at a Glance. Tokyo: Academia Music Ltd.

Trehub, S. E., Becker, J., and Morley, I. (2015). Cross-cultural perspectives on music and musicality. Philosophical Transactions of the Royal Society B: Biological Sciences. Royal Society of London. https://doi.org/10.1098/rstb.2014.0096

Woods, D.L., Stecker G.C., Rinne T., Herron T.J., Cate A.D., et al. (2009). Functional Maps of Human Auditory Cortex: Effects of Acoustic Features and Attention. PLoS ONE. 4(4): e5183. https://doi.org/10.1371/journal.pone.0005183

Wong, P. C., Roy, A. K., and Margulis, E. H. (2009). Bimusicalism: The implicit dual enculturation of cognitive and affective systems. Music Perception, 27(2), 81-88. https://doi.org/10.1525/mp.2009.27.2.81

Yue, Q., Martin, R. C., Fischer-Baum, S., Ramos-Nuñez, A. I., Ye, F., and Deem, M. W. (2017). Brain modularity mediates the relation between task complexity and performance. Journal of Cognitive Neuroscience. 29, 1532–1546. https://doi.org/10.1162/jocn_a_01142

Zatorre, R. J., and Salimpoor, V. N. (2013). From perception to pleasure: music and its neural substrates. Proceedings of the National Academy of Sciences USA. 110, 10430–10437. https://doi.org/10.1073/pnas.1301228110

Zwicker, E. (1961). Subdivision of the audible frequency range into critical bands (Frequenzgruppen). The Journal of the Acoustical Society of America, 33(2), 248-248. https://doi.org/10.1121/1.1908630